\documentclass[final,1p,times,authoryear]{elsarticle}

\newcommand\apj{ApJ}
\newcommand\apjl{ApJ}
\newcommand\apjs{ApJS}
\newcommand\aj{AJ}
\newcommand\mnras{MNRAS}
\newcommand\aap{A\&A}
\newcommand\araa{ARA\&A}

\def\la{\ \raise0.3ex\hbox{$<$}\kern-0.75em{\lower0.65ex\hbox{$\sim$}}\ }
\def\ga{\ \raise0.3ex\hbox{$>$}\kern-0.75em{\lower0.65ex\hbox{$\sim$}}\ }

\journal{New Astronomy}

\begin{document}

\begin{frontmatter}

\title{Magnetic Field and Faraday Rotation Measure in the Turbulent Warm Ionized Medium}

\author[l1]{Qingwen Wu}\ead{qwwu@mail.hust.edu.cn}
\author[l2]{Jongsoo Kim}\ead{jskim@kasi.re.kr}
\author[l3]{Dongsu Ryu\corref{cor1}}\ead{ryu@canopus.cnu.ac.kr}
\address[l1]{School of Physics, Huazhong University of Science and Technology, Wuhan 430074, China}
\address[l2]{Korea Astronomy and Space Science Institute, Daejeon 305-348, Korea}
\address[l3]{Department of Astronomy and Space Science, Chungnam National University, Daejeon 305-764, Korea}
\cortext[cor1]{Corresponding author. Tel.: +82 42 821 5466}

\begin{abstract}

\citet{wu09} suggested an empirical relation between the magnetic field strength along the line of sight (LOS) and the dispersion of Faraday rotation measure (RM) distribution in turbulent media with root-mean-square sonic Mach number $M_s \simeq 1$. In this paper, we extend the work by incorporating the Mach number dependence. Media with $0.5 \la M_s \la 2$ are considered to cover the Mach number range of the warm ionized medium (WIM) of our Galaxy. Three-dimensional, magnetohydrodynamic isothermal turbulence simulations with solenoidal forcing are used. We suggest a new relation among the LOS magnetic field strength, the dispersion of RM distribution, and the Mach number, which approximately represents the relation for Alfv\'enic perturbations. In addition, we suggest a relation between the Mach number and the dispersion of log-normal distribution of emission measure (EM), which is basically the relation for the Mach number and the density dispersion. The relations could be used for a quick and rough estimation of the LOS magnetic field strength in the turbulent WIM.
\end{abstract}


\end{frontmatter}

\section{Introduction}

The ionized gas in the Galaxy has been traditionally associated with bright regions surrounding hot stars, called Str$\ddot{\rm o}$mgren spheres or classical H II regions. Most H II regions, however, are found only at low Galactic latitudes with a scale height of $\sim 40 - 70$ pc, which is much smaller than that of the diffuse ionized gas (DIG) or the warm ionized medium (WIM) \citep[e.g.,][]{go01}. In fact, the classical H II regions contain only $\sim 10 \%$ of ionized hydrogen in the Galaxy, and the remaining $90 \%$ resides in warm ($\bar{T} \sim 8000\rm\ K$) and diffuse ($\bar{n}\sim 0.03 \ \rm cm^{-3}$) regions. The WIM occupies approximately 20 - 30 \% of the volume of $\sim 2\ \rm kpc$-thick, plane-parallel layer of our Galaxy \citep[e.g.,][]{reyn91,ha99,ga08,hi08}. The Wisconsin ${\rm H}\alpha$ Mapper (WHAM) northern sky survey has provided information on the distribution, kinematics, and other physical properties of the WIM \citep{ha03}. The diffuse WIM is now recognized as one of the major components of the interstellar media (ISM) in both our Galaxy and external galaxies \citep[e.g.,][]{ra90,reyn91,cr01,mv03} (see also \citet{ha09} for a review).

The properties of the WIM have been revealed through observations of dispersion measure (DM)
\begin{equation}
{\rm DM} = \int n_e ds
\end{equation}
and emission measure (EM)
\begin{equation}
{\rm EM} = \int n_e^2 ds,
\end{equation}
where $n_e$ is the electron density and $s$ is the length along the line of sight (LOS). For instance, EM/DM and DM$^{2}$/EM give estimations of the electron density and the extent of the WIM, respectively \citep[e.g.,][]{reyn91,be06,hi08}. In addition, the widths of probability density functions (PDFs) of DM and EM are known to be related to the root-mean-square (rms) sonic Mach number, $M_s$, of the WIM \citep[e.g.,][]{bf08,hi08}.

The WIM is most likely to be in a state of turbulence, considering a very high Reynolds number inferred from observed velocities and expected viscosity \citep{be99}. Evidences for turbulence include the log-normal distributions of 1) EMs from the H$\alpha$ intensity in the Wisconsin H$\alpha$ Mapper \citep{ha99,ha03} survey \citep{hi08}, 2) EMs from the H$\alpha$ intensity in external galaxies \citep[e.g.,][]{ta07,se09}, and 3) column densities of the DIG in our Galaxy \citep{bf08}. The log-normality can be explained naturally by turbulent distribution of gas \citep[e.g.,][]{vs94,es04,kri07,fe08,fe10}. The best evidence for turbulence comes from the power spectrum presented in \citet{ar95}. It is a composite power spectrum of electron density collected from observations of velocity fluctuations of the interstellar gas, rotation measures (RMs), DMs, interstellar scintillations, and others. The spectrum covers a huge range of $\sim 10^{10} - 10^{20}$ cm. The whole range of the spectrum is approximately fitted to the power spectrum of Kolmogorov turbulence with slope $-5/3$.

The WIM is permeated with magnetic fields. Measuring Faraday rotation of polarized radio emissions against background sources has been the most popular method for exploring magnetic fields there. The rotation of the plane of linearly polarized light is due to the birefringence of magneto-ionic medium. The angle of rotation is given by
\begin{equation}
\psi = {\rm RM}\ \lambda^2,
\end{equation}
where RM is the rotation measure defined by
\begin{equation}
{\rm RM} = 0.81 \int n_e B_\parallel ds.
\end{equation}
Here, $\lambda$ is the wavelength of polarized light and $B_{\parallel}$ is the LOS magnetic field strength.  The units of RM, $n_e$, $B_\parallel$ and $s$ are rad m$^{-2}$, cm$^{-3}$, $\mu$G, and pc, respectively. RM/DM gives an estimation of the LOS magnetic field strength, weighted by the electron density,
\begin{equation}
\left< B_{\parallel} \right> = \int n_{e} B_{\parallel} ds\ {\Bigl /} \int n_e ds.
\end{equation}
The method, for instance, has been used to reconstruct the large-scale magnetic field in our Galaxy by many authors \citep[e.g.,][]{han98,id99,fr01,han06,be07}.

\citet{hav03,hav04} obtained the distributions of RMs along contiguous LOSs in the constellations of Auriga and Horologium in multi-frequency polarimetric observations with the Westerbork Synthesis Radio Telescope. While the peak in the frequency distribution of those RMs reflects the regular component of magnetic field, $B_0$, the spread should measure the turbulent component. This means that if a distribution of RM is observed, its spread provides another way to quantify the magnetic field in turbulent ionized media such as the WIM. Recently, using numerical simulations, \citet{wu09} found that in turbulent media with $M_s \simeq 1$, the width of the distribution of RM/$\overline{\rm RM}$ ($\overline{\rm RM}$ is the average value of RMs) is rather tightly related to the strength of the regular field along the LOS, $B_{0\parallel}$. They suggested an empirical formula, which can be used to estimate $B_{0\parallel}$ when the distribution of RM/$\overline{\rm RM}$ is available.

However, the Mach number is not necessarily unity in the WIM. The turbulent velocity dispersion has a range of values $v_{\rm turb} \sim 10 - 30$ km s$^{-1}$ \citep{trh99}, and the temperature also has a range values $T \sim 6000 - 12000$ K \citep[e.g.,][]{re99}. So in the WIM, the turbulent flow motions, although not always having $M_s \simeq 1$, are expected to be somewhere between mildly subsonic and mildly supersonic. As a matter of fact, several groups have suggested that the turbulent WIM has $M_s \simeq 1 - 2.5$ \citep[e.g.,][]{hi08,bu09,ga11,bu12}.

In this paper, we extend the work of \citet{wu09} by incorporating the Mach number dependence. Specifically, we consider turbulent media with $M_s \simeq 0.5$ and 2 along with $M_s \simeq 1$. In Section 2, we outline our simulations. In Section 3, we present a new relation among $B_{0\parallel}$, RM/$\overline{\rm RM}$, and $M_s$. We also suggest a relation between $M_s$ and the dispersion of log-normal distribution of EM, which can be used to estimate the Mach number. Summary follows in Section 4.

\section{Simulations}

We performed three-dimensional simulations using a code based on the total variation diminishing (TVD) scheme \citep{kim99}, by solving the following set of equations for isothermal, compressible magnetohydrodynamics (MHDs)
\begin{equation}
\frac{\partial\rho}{\partial t} + \mbox{\boldmath$\nabla\cdot$}\left(\rho\mbox{\boldmath$v$}\right) = 0,
\end{equation}
\begin{equation}
\rho\left(\frac{\partial\mbox{\boldmath$v$}}{\partial t} + \mbox{\boldmath$v\cdot\nabla v$}\right) + c_s^2\mbox{\boldmath$\nabla$}\rho - \frac{1}{4\pi}\left(\mbox{\boldmath$\nabla\times B$}\right) \mbox{\boldmath$\times B$} = \rho {\mbox{\boldmath$f$}},
\end{equation}
\begin{equation}
\frac{\partial\mbox{\boldmath$B$}}{\partial t} -\mbox{\boldmath$\nabla\times$}\left(\mbox{\boldmath$v\times B$}\right)
= 0,
\end{equation}
\begin{equation}
\mbox{\boldmath$\nabla\cdot B$} = 0,
\end{equation}
where $c_s$ is the isothermal sound speed. Turbulence was driven by imposing a solenoidal (incompressible) forcing, \mbox{\boldmath$f$}. We followed the recipes of \citet{st99} and \citet{ml99} for \mbox{\boldmath$f$}. Perturbations, satisfying \mbox{\boldmath$k$}$\cdot$\mbox{\boldmath$f$}$_k = 0$, were drawn from a Gaussian random field in the Fourier space of the wavevector \mbox{\boldmath$k$}. The magnitude was determined by the top-hat power distribution in a narrow wave-number range of $(2\pi/L)\leq k \leq 2 (2\pi/L)$, where $L$ is the computational box size. The perturbations were converted to quantities in the real space by Fourier transform, and then added into the computational domain at every $\delta t_f = 0.001 L/c_{\rm s}$. In contrast to the original recipes, we used a different seed number for realization of perturbations at every $\delta t_f$. The amplitude of the forcing was fixed in such a way that the resulting input rate of the kinetic energy is a constant. Initially a static, uniformly magnetized medium with density $\rho_0$ and magnetic field $B_0$ along the $x$-direction was assumed. A periodic computational box with $512^3$ grid zones was used.

There are two parameters in the problem, the initial plasma beta, $\beta_0 = \rho_0c_s^2/(B_0^2/(8\pi))$, and the rms sound Mach number at the saturated stage of turbulence, $M_s = v_{\rm rms}/ c_s$. To cover the ranges of values expected in the WIM of our galaxy, we included the cases with $\beta_0 =$ 0.1, 1, 10 and $M_s = 0.5$, 1, 2. The values of $\beta_0$ were set by the initial condition. For $M_s$, the amplitude of forcing was set by trial and error, so that $M_s$ became close to the predefined value at the saturated stage. We present nine simulations in this paper.

The initial magnetic field strength is related to the dimensionless parameter, $\beta_0$, by
\begin{equation}
B_0 = 1.3 \left( 1 \over \beta_0 \right)^{1/2} \left( T \over 8000\ {\rm K} \right)^{1/2} \left( n_e \over 0.03\ {\rm cm}^{-3} \right)^{1/2}\ \mu{\rm G},
\end{equation}
assuming that hydrogen is completely ionized, helium is neutral, and the number ratio of hydrogen to helium is 10. If we take 8000 K and 0.03 cm$^{-3}$ as the representative values of temperature and electron density in the WIM, the initial magnetic field strength in our simulations corresponds to 4.1, 1.3, 0.41 $\mu$G for $\beta_0$ = 0.1, 1 and 10, respectively. It covers the range of the regular magnetic field strength in our Galaxy \citep[e.g.,][and references therein]{han06}.

Figure 1 shows the evolution of $M_s$ as a function of time in units of $t_{\rm turb} \equiv L/(2 M_s c_s)$. Note that $t_{\rm turb}$ is the turbulent turnover time in our simulations. The rms flow speed, $v_{\rm rms}$, and so $M_s$ initially increase and then saturate. Saturation is reached around $\sim 2\ t_{\rm turb}$, as previously shown \citep[e.g.,][]{fe09}. We ran simulations up to $4\ t_{\rm turb}$. We then took 11 snapshots (black dots in Figure 1) at the saturated stage and used them to calculating the quantities in following sections. The errors below are the standard deviations of the 11 snapshot data.

The magnetic field strength estimated with RM using Equation (5) would be unbiased, only if the correlation between $B_{\parallel}$ and $n_{e}$ is null or weak, as noted in, e.g., \citet{be03}. Both observations \citep{crut99,pn99} and numerical simulations \citep[e.g.,][]{os01,pas03,bk05,ml05,bu09} have indicated that the correlation would be negative and positive in subsonic and supersonic turbulent flows, respectively. We calculated the correlation coefficients for our simulation data,
\begin{equation}
r(B,\rho)=
\frac{{\Sigma}_{i,j,k} (B_{i,j,k}-\bar B)(\rho_{i,j,k}-\bar \rho)}
{\left[\Sigma_{i,j,k}(B_{i,j,k}-\bar B)^2\right]^{1/2}
\left[\Sigma_{i,j,k}(\rho_{i,j,k}-\bar \rho)^2\right]^{1/2}},
\end{equation}
where $\bar B$ and $\bar \rho$ are the average values of $B$ and $\rho$. Table 1 shows the resulting correlation coefficients for nine simulations. As in \citet{wu09}, we have small values, $|r| \la 0.1$, except for the case of $M_s = 2$ and $\beta_0 = 10$; even in the case, we have $r = 0.23 \pm 0.03$. The weak correlation means that the RM field strength in Equation (5) should correctly represent the true magnetic field strength, as further discussed in \citet{wu09}. The correlation in our simulations looks consistent to those of previous works. For instance, we have $r = -0.1 \pm 0.05$ for $\beta_0 = 1$ and $M_s =1$, and \citet{bu09} reported a negative correlation for $\beta_0 = 2$ and $M_s =0.7$.

We note that the correlation coefficients would depend not only on $\beta_0$ and $M_s$, but also on the nature of forcing. It is well known that the properties of turbulence, such as the density PDF and power spectrum, depend on whether the forcing is solenoidal or compressible \citep[see, e.g.,][]{fe08,fe09,fe10}. Recently, for instance, \citet{fe13} reported a study of supersonic turbulence with solenoidal and compressive drivings where the density PDF was analyzed in details. In general, compressible forcing results in stronger compression and so larger standard deviation of $\ln \rho$, $\sigma_{\ln \rho}$, than the solenoidal forcing. It tells that the correlation coefficient is expected to be different in turbulences driven by solenoidal and compressible forcings, and the results presented in the next Section could be affected by the nature of forcing. In this paper, we consider only the turbulence with solenoidal forcing, leaving the effects of the nature of forcing as a future study.

\section{Results}

\subsection{Relation among $B_{0\parallel}$, $W_{\rm FWHM(RM/\overline{RM})}$, and $M_s$}

\citet{wu09} demonstrated that the frequency distribution of RMs in turbulence simulation is well fitted to the Gaussian for the case of $M \simeq 1$. Furthermore, $B_{0\parallel}$ is shown to be anti-correlated with the full width at half maximum (FWHM) of the frequency distribution of RM/$\overline{\rm RM}$, $W_{\rm FWHM(RM/\overline{RM})}$, as noted in Introduction. From a physical point of view, the broadening of the width of RM distribution is caused by fluctuating gas and magnetic field. So it is easily expected that $W_{\rm FWHM(RM/\overline{RM})}$ would also depend on $M_s$. Here, we further explore the possible relation among $B_{0\parallel}$, $W_{\rm FWHM(RM/\overline{RM})}$, and $M_s$ based on the new simulations described in the previous section.

To obtain the relation among $B_{0\parallel}$, $W_{\rm FWHM(RM/\overline{RM})}$, and $M_s$, we took five viewing angles, $\theta = 0^\circ,\ 27^\circ,\ 45^\circ,\ 63^\circ,$ and $83^\circ$, where $\theta$ is the angle between the regular magnetic field, \mbox{\boldmath$B$}$_0$, and the LOS ($B_{0\parallel} \equiv B_0 \cos\theta$). We first calculated RM for the $\theta$'s in our nine simulations. Table 2 show $\rm \overline{RM}$, normalized with RM due to $B_{0\parallel}$, ${\rm RM_0} = 0.81 n_0 B_{0\parallel} L$, where $n_0$ is the number density of the background medium. $\rm \overline{RM}/RM_0$ is close to unity indicating that $\rm \overline{RM}$ is caused mostly by $B_{0\parallel}$, except for $\theta = 83^\circ$ where the contribution of perturbed $B_{0\perp} (\equiv B_0 \sin\theta)$ is more important. We then calculated the frequency distribution of RM/$\overline{\rm RM}$, $f_{\rm RM/\overline{RM}}$, and $W_{\rm FWHM(RM/\overline{RM})}$. For the Gaussian distribution, the FWHM is related to the standard deviation, $\sigma$, as $W_{\rm FWHM} = 2 \sqrt{2 \ln 2}\ \sigma$. The resulting 45 FWHMs are listed in Table 3. Figure 2 shows $W_{\rm FWHM(RM/\overline{RM})}$ versus the LOS regular field strength, $B_{0\parallel} \equiv B_0 \cos\theta$, for the 45 FWHMs. Here, the magnetic field strength is for the representative values, $T = 8000$ K and $n_e = 0.03$ cm$^{-3}$. $W_{\rm FWHM(RM/\overline{RM})}$ is larger for larger $\beta_0$ and for larger $\theta$, as already pointed in \citet{wu09}. So $W_{\rm FWHM(RM/\overline{RM})}$ is anti-correlated with $B_{0\parallel}$. For each $M_s$, we fitted the relation between $B_{0\parallel}$ and $W_{\rm FWHM(RM/\overline{RM})}$ with dashed lines. The broadening of the width of RM distribution is due to the fluctuating magnetic field and electron density. So as expected, $W_{\rm FWHM(RM/\overline{RM})}$ is larger for larger $M_s$.

We got the relation among $B_{0\parallel}$, $W_{\rm FWHM(RM/\overline{RM})}$, and $M_s$ by fitting the 45 FWHMs in Table 3 to a single formula, $B_{0\parallel}=c_1\times M_s^{c_2}/W_{\rm FWHM(RM/\overline{RM})}^{c_3}$, where $c_1,\ c_2,$ and $c_3$ are fitting parameters. The best fit we found\footnote{We note that the frequency distribution of RM/$\overline{\rm RM}$, $f_{\rm RM/\overline{RM}}$, was used to calculate $W_{\rm FWHM(RM/\overline{RM})}$ in this paper, while the log of it, $\log_{10}(f_{\rm RM/\overline{RM}})$, was used in \citet{wu09}. It is because published observations mostly provide $f_{\rm RM/\overline{RM}}$ (see the next section), and so it is easier to measure the FWHM of $f_{\rm RM/\overline{RM}}$. We found that the FWHM of $f_{\rm RM/\overline{RM}}$ is $\sim 2 - 3.5$ times smaller than the FWHM of $\log_{10}(f_{\rm RM/\overline{RM}})$. With the FWHM of $\log_{10}(f_{\rm RM/\overline{RM}})$, we would obtain the fitted relation that is consistent with that of \citet[][Equation (3)]{wu09}.} is
\begin{equation}
B_{0\parallel} = (0.65\pm0.02) \times \frac{M_s^{1.19\pm0.07}}{W_{\rm FWHM(RM/\overline{RM})}^{1.31\pm0.04}}\ \mu{\rm G},
\end{equation}
where the errors indicate the fitting uncertainty. Again, $B_{0\parallel}$ is for $T = 8000$ K and $n_e = 0.03$ cm$^{-3}$, and scales as $(T/8000\ {\rm K})^{1/2} (n_{\rm e}/0.03\ {\rm cm}^{-3})^{1/2}$ for other values of $T$ and $n_{\rm e}$ (see Equation (10)). Figure 3 shows $B_{0\parallel}$ as a function of $0.65 \times M_s^{1.19}/ W_{\rm FWHM(RM/\overline{RM})}^{1.31}$ for FWHMs in Figure 2 along with the fitting, demonstrating the goodness of the fitting. The empirical relation in Equation (12) would provide a handy way to quantify the LOS regular field strength in regions where the Mach number and the RM distribution have been obtained. It is interesting to see that the relation is applied even to the case of fairly large viewing angle, $\theta = 83^\circ$. However, it is clear that the relation should break down if $B_{0\parallel} \sim 0$ and so ${\rm\overline{RM}} \sim 0$. So it can be applied only to regions with ${\rm\overline{RM}}$ not too small.

The above relation can be approximately reproduced from $\delta B / B_0 \approx v / c_A$ for Alfv\'enic perturbations, where $c_A$ is the Alfv\'en speed. In the case that the correlation between $B$ and $\rho$ is weak (see Section 2), roughly $\delta B \propto \delta {\rm RM}$, so $\delta B / B_0 \propto W_{\rm FWHM(RM/\overline{RM})}$. Also, $v = c_s M_s \propto \sqrt{T} M_s$ and $c_A = B_0 / \sqrt{4\pi\rho}$. Combining these, we have $B_0 \propto \sqrt{T\rho}\ M_s / W_{\rm FWHM(RM/\overline{RM})}$. Of course, the perturbations in our simulations are not totally Alfv\'enic and the correlation between $B$ and $\rho$ is not completely null. So we got the relation with the exponents which are somewhat different from unity.

\subsection{Relation between $M_s$ and  $W_{\rm FWHM[\log_{10}({\rm EM})]}$}

To derive $B_{0\parallel}$ using Equation (12), it requires us to know not only the RM distribution, but also the Mach number. It would be handy if we have an independent way to determine the Mach number. From numerical simulations for isothermal, hydrodynamic or MHD turbulence, it is known that the density PDF is approximately fitted to the log-normal distribution, and its standard deviation, $\sigma_{\ln\rho}$, increases as the rms Mach number of turbulent flows increases \citep{vs94,pa97,np99,os99,os01,cho03,es04,kri07,ko07,fe10}. It has been shown that $\sigma_{\ln\rho}$ could be related to $M_s$ as $\sigma_{\ln\rho}^2 = \ln(1+b^2M_s^2)$ with $b \sim 0.3$ for turbulence with solenoidal forcing and $b \sim 1$ for turbulence with compressible forcing, regardless of the presence of the magnetic field \citep[see, e.g.,][]{pa97,kri07,fe10}. Observationally, however, it is not trivial to get the distribution of volume density; it is easier to measure the distribution of column density (DM) or EM. It has been argued that the column density and EM follow the log-normal distribution too \citep[e.g.,][]{os01,ko07,hi08,fe10,fe13}.

Here, we look for a relation between the dispersion of EM distribution and the Mach number, which may be used to estimate $M_s$. We obtained the frequency distribution of $\log_{10}\rm (EM)$ for five angles, $\theta = 0^\circ,\ 27^\circ,\ 45^\circ,\ 63^\circ,\ 90^\circ$, in our nine simulations. The distribution is fitted to the Gaussian. We then calculated the FWHM, $W_{\rm FWHM[\log_{10}({\rm EM})]}$, of the distribution. Table 4 lists the resulting 45 FWHMs. Figure 4 shows $W_{\rm FWHM[\log_{10}({\rm EM})]}$ versus $M_s$ for the 45 FWHMs. $W_{\rm FWHM[\log_{10}({\rm EM})]}$ is most sensitive to $M_s$, while its dependence on $\beta_0$ and $\theta$ is weaker. The average values of $W_{\rm FWHM[\log_{10}({\rm EM})]}$ are $0.17\pm0.04$, $0.33\pm0.06$, $0.58\pm0.09$ for $M_s$=0.5, 1, and 2, respectively, where the average and standard deviation are taken for 165 data (5 $\theta'\rm s\ \times\ 3 \beta_0'\rm s\ \times$ 11 snapshots). They are also shown in Figure 4 too. Our result is roughly consistent with that of \citet{hi08}, where the width was from simulations with $256^{3}$ grid zones. Filled hexagons in Figure 4 plot $W_{\rm FWHM}$ (converted from $\sigma$) in their Table 5. \citet{hi08} also argued that the dependence of $W_{\rm FWHM[\log_{10}({\rm EM})]}$ on magnetic field is weak. We fitted $M_s$ versus $W_{\rm FWHM[\log_{10}({\rm EM})]}$ to a linear function, and the best fit we got is
\begin{equation}
M_s = 3.60(\pm0.84) W_{\rm FWHM[\log_{10}({\rm EM})]} - 0.13(\pm0.26).
\end{equation}

The above can be reproduced from $\sigma_{\ln\rho}^2 = \ln (1 + b^2 M_s^2)$. With $\sigma_{\ln\rho} \approx b M_s$ for $M_s \sim 1$ ($b \sim 0.3$ for solenoidal forcing) and $\sigma_{\ln\rho} \propto W_{\rm FWHM[\log_{10}({\rm EM})]}$ for $\sigma_{\ln\rho} < 1$, we have $M_s \propto W_{\rm FWHM[\log_{10}({\rm EM})]}$, which is close to our fitted relation. As $\sigma_{\ln\rho}^2 = \ln (1 + b^2 M_s^2)$, our relation should depend on forcing; it is applicable only to the case of solenoidal forcing.

\subsection{Correlation between RM and EM}

We also checked the correlation between RM and EM. Both quantities involve the electron density along the LOS, and so a positive correlation is expected. We calculated the correlation coefficients between RM and $\sqrt {\rm EM}$ with two-dimensional, spatial distributions, for five viewing angles, $\theta = 0^\circ,\ 27^\circ,\ 45^\circ,\ 63^\circ,\ 90^\circ$, in nine simulations using a formula similar to that in Equation (11). The resulting coefficients are listed in Table 5. For small $\beta_0$'s and $\theta$'s, that is, for the cases with sufficiently large $B_{0\parallel}$'s, there is a quite strong, positive correlation, as expected. The correlation, however, is weaker for larger $\beta_0$'s and $\theta$'s. Our result shows that the correlation is very week and can even become negative for $\theta = 90^\circ$.

\section{Summary}

\citet{wu09} found a relation between the magnetic field strength along the LOS, $B_{0\parallel}$, and the FWHM of the frequency distribution of RM/$\overline{\rm RM}$, $W_{\rm FWHM(RM/\overline{RM})}$, for turbulent media of $M_s \simeq 1$. But the Mach number in the WIM is not necessarily unity. In this paper, we incorporated the Mach number dependence in the relation. For it, we performed three-dimensional simulations of isothermal, compressible MHD turbulence for $M_s$ = 0.5, 1, 2 and $\beta_0$ = 0.1, 1, 10. The parameters were chosen to cover the range of values expected in the WIM of our Galaxy.

- From the frequency distribution of RM/$\overline{\rm RM}$ which is well fitted with the Gaussian, we calculated $W_{\rm FWHM(RM/\overline{RM})}$. We suggest a relation among $B_{0\parallel}$, $W_{\rm FWHM(RM/\overline{RM})}$, and $M_s$, which is shown in Equation (12).

- The frequency distribution of EM is well fitted with the log-normal distribution. We calculated the FWHM of the frequency distribution of $\log_{10}\rm (EM)$, $W_{\rm FWHM[\log_{10}({\rm EM})]}$. We suggest a relation between $M_s$ and  $W_{\rm FWHM[\log_{10}({\rm EM})]}$, which is shown in Equation (13).

The relation in Equation (12) would provide a handy way for a quick and rough estimation of $B_{0\parallel}$ in the turbulent WIM regions where observations of RM distribution are available and $M_s$ is known. In the case that $M_s$ is unknown, the relation in Equation (13) could be used for an estimation of $M_s$, if observations of EM distribution are available.

\section*{Acknowledgments}

The work of QW was supported by the NSFC (grants 11143001, 11103003, 11133005), the National Basic Research Program of China (2009CB824800), the Doctoral Program of Higher Education (20110142120037), and the Fundamental Research Funds for the Central Universities (HUST: 2011TS159). The work of JK was supported by National Research Foundation of Korea through grant K20901001400-10B1300-07510. The work of DR was supported by a research fund of Chungnam National University. Numerical simulations were performed by using a high performance computing cluster at the Korea Astronomy and Space Science Institute.

\clearpage

\begin{table}
\begin{center}
\begin{tabular}{c|ccc}\hline
\multicolumn{4}{c}{$r(B,\rho)$}\\ \hline\hline
                & $\beta_0$=0.1  & $\beta_0$=1     & $\beta_0$=10   \\ \hline
$M_s\simeq0.5$  & $-0.06\pm0.05$ & $-0.13\pm0.04$  & $-0.07\pm0.04$ \\ \hline
$M_s\simeq1$    & $-0.05\pm0.05$ & $-0.10\pm0.05$  & $0.05\pm0.03$  \\ \hline
$M_s\simeq2$    & $ 0.00\pm0.04$ & $0.10\pm0.03$   & $0.23\pm0.03$  \\ \hline
\end{tabular}
\caption{Correlation coefficients between $B$ and $\rho$ calculated from the Eq. 11. The error is the standard deviation of 11 snapshot data.}
\end{center}
\end{table}

\clearpage

\begin{table}
\begin{center}
\begin{tabular}{c|ccc}\hline
\multicolumn{4}{c}{$\rm \overline{RM}/RM_0$}\\ \hline\hline
                   & $\beta_{0}$=0.1 & $\beta_{0}$=1 & $\beta_{0}$=10\\
\hline
  &     & $M_s \simeq 0.5$  &   \\
  $\theta=0^\circ$     & $1.00\pm0.001$   & $0.99\pm0.001$   & $0.99\pm0.005$   \\
  $\theta=27^\circ$    & $1.00\pm0.001$   & $1.00\pm0.001$   & $0.99\pm0.006$   \\
  $\theta=45^\circ$    & $1.00\pm0.001$   & $1.00\pm0.002$   & $0.99\pm0.007$   \\
  $\theta=63^\circ$    & $1.00\pm0.001$   & $1.00\pm0.003$   & $0.99\pm0.01$   \\
  $\theta=83^\circ$    & $0.02\pm0.0001$  & $0.02\pm0.0002$  & 0.02$\pm0.0005$   \\
\hline
  &     & $M_s \simeq 1$    &   \\
  $\theta=0^\circ$     & $1.00\pm0.001$   & $0.98\pm0.004$    & $0.96\pm0.01$   \\
  $\theta=27^\circ$    & $1.00\pm0.002$   & $0.98\pm0.004$    & $0.97\pm0.01$   \\
  $\theta=45^\circ$    & $1.00\pm0.002$   & $0.98\pm0.007$    & $0.97\pm0.02$   \\
  $\theta=63^\circ$    & $1.00\pm0.004$   & $0.97\pm0.01$     & $0.98\pm0.03$   \\
  $\theta=83^\circ$    & $0.02\pm0.0002$  & $0.06\pm0.0006$    & $0.02\pm0.002$   \\
\hline
  &     & $M_s \simeq 2$    &   \\
  $\theta=0^\circ$     & $0.99\pm0.004$   & $0.97\pm0.01$   & $1.02\pm0.04$   \\
  $\theta=27^\circ$    & $0.99\pm0.005$   & $0.97\pm0.02$   & $0.99\pm0.03$   \\
  $\theta=45^\circ$    & $0.99\pm0.009$   & $0.96\pm0.03$   & $0.97\pm0.04$   \\
  $\theta=63^\circ$    & $0.99\pm0.02$    & $0.96\pm0.05$   & $0.91\pm0.06$   \\
  $\theta=83^\circ$    & $0.02\pm0.001$   & $0.01\pm0.003$  & $0.01\pm0.003$   \\
\hline
\end{tabular}
\caption{Average of normalized RM. Here, ${\rm RM_0} = 0.81 n_0 B_{0\parallel} L$. The error is the standard deviation of 11 snapshot data.}
\end{center}
\end{table}

\clearpage

\begin{table}
\begin{center}
\begin{tabular}{c|ccc}\hline
\multicolumn{4}{c}{$W_{\rm FWHM(RM/\overline{RM})}$}\\ \hline\hline
                   & $\beta_{0}$=0.1 & $\beta_{0}$=1 & $\beta_{0}$=10\\
\hline
  &     & $M_s \simeq 0.5$  &   \\
  $\theta=0^\circ$     & $0.11\pm0.01$   & $0.26\pm0.02$   & $0.52\pm0.03$   \\
  $\theta=27^\circ$    & $0.16\pm0.01$   & $0.27\pm0.01$   & $0.56\pm0.05$   \\
  $\theta=45^\circ$    & $0.20\pm0.01$   & $0.31\pm0.03$   & $0.62\pm0.04$   \\
  $\theta=63^\circ$    & $0.22\pm0.03$   & $0.41\pm0.06$   & $1.12\pm0.13$   \\
  $\theta=83^\circ$    & $0.53\pm0.06$   & $1.40\pm0.14$   & $4.49\pm0.35$   \\
\hline
  &     & $M_s \simeq 1$    &   \\
  $\theta=0^\circ$     & $0.26\pm0.01$   & $0.57\pm0.03$    & $1.01\pm0.11$   \\
  $\theta=27^\circ$    & $0.30\pm0.04$   & $0.62\pm0.03$    & $1.12\pm0.12$   \\
  $\theta=45^\circ$    & $0.37\pm0.05$   & $0.68\pm0.06$    & $1.23\pm0.09$   \\
  $\theta=63^\circ$    & $0.37\pm0.03$   & $0.79\pm0.05$    & $2.07\pm0.10$   \\
  $\theta=83^\circ$    & $0.99\pm0.10$   & $3.17\pm0.47$    & $7.07\pm0.84$   \\
\hline
  &     & $M_s \simeq 2$    &   \\
  $\theta=0^\circ$     & $0.52\pm0.02$   & $0.97\pm0.08$    & $1.58\pm0.04$   \\
  $\theta=27^\circ$    & $0.62\pm0.04$   & $1.01\pm0.06$    & $1.71\pm0.05$   \\
  $\theta=45^\circ$    & $0.64\pm0.03$   & $1.11\pm0.06$    & $1.91\pm0.08$   \\
  $\theta=63^\circ$    & $0.67\pm0.04$   & $1.42\pm0.14$    & $3.67\pm0.42$   \\
  $\theta=83^\circ$    & $1.92\pm0.16$   & $5.71\pm0.92$    & $19.21\pm4.02$  \\
\hline
\end{tabular}
\caption{Full width at half maximum of the frequency distribution of $\rm RM/\overline{RM}$. The error is the standard deviation of 11 snapshot data.}
\end{center}
\end{table}

\clearpage

\begin{table}
\begin{center}
\begin{tabular}{c|ccc}\hline
\multicolumn{4}{c}{$W_{\rm FWHM[\log_{10}({\rm EM})]}$}\\ \hline\hline
                   & $\beta_{0}$=0.1 & $\beta_{0}$=1.0 & $\beta_{0}$=10.0\\
\hline
  &     & $M_s \simeq 0.5$  &    \\
  $\theta=0^\circ$     & $0.06\pm0.01$   & $0.16\pm0.03$   & $0.20\pm0.02$   \\
  $\theta=27^\circ$    & $0.16\pm0.01$   & $0.18\pm0.01$   & $0.19\pm0.01$   \\
  $\theta=45^\circ$    & $0.17\pm0.02$   & $0.17\pm0.02$   & $0.19\pm0.02$   \\
  $\theta=63^\circ$    & $0.17\pm0.01$   & $0.19\pm0.01$   & $0.19\pm0.02$   \\
  $\theta=90^\circ$    & $0.18\pm0.02$   & $0.21\pm0.02$   & $0.18\pm0.02$   \\
\hline
  &     & $M_s \simeq 1$    &    \\
  $\theta=0^\circ$     & $0.17\pm0.01$   & $0.33\pm0.02$   & $0.38\pm0.03$   \\
  $\theta=27^\circ$    & $0.29\pm0.02$   & $0.34\pm0.03$   & $0.38\pm0.03$   \\
  $\theta=45^\circ$    & $0.29\pm0.03$   & $0.34\pm0.04$   & $0.37\pm0.04$   \\
  $\theta=63^\circ$    & $0.30\pm0.02$   & $0.37\pm0.05$   & $0.36\pm0.04$   \\
  $\theta=90^\circ$    & $0.33\pm0.03$   & $0.38\pm0.03$   & $0.37\pm0.03$   \\
\hline
  &     & $M_s \simeq 2$    &    \\
   $\theta=0^\circ$    & $0.41\pm0.02$   & $0.63\pm0.06$   & $0.65\pm0.06$   \\
  $\theta=27^\circ$    & $0.47\pm0.03$   & $0.60\pm0.03$   & $0.64\pm0.05$   \\
  $\theta=45^\circ$    & $0.46\pm0.03$   & $0.57\pm0.06$   & $0.60\pm0.07$   \\
  $\theta=63^\circ$    & $0.55\pm0.06$   & $0.61\pm0.07$   & $0.64\pm0.07$   \\
  $\theta=90^\circ$    & $0.59\pm0.05$   & $0.64\pm0.06$   & $0.65\pm0.07$   \\
\hline
\end{tabular}
\caption{Full width at half maximum of the frequency distribution of $\log_{10}\rm (EM)$. The error is the standard deviation of 11 snapshot data.}
\end{center}
\end{table}

\clearpage

\begin{table}
\begin{center}
\begin{tabular}{l|ccc}\hline
\multicolumn{4}{c}{$r({\rm RM},\sqrt{\rm EM})$}\\ \hline\hline
                   & $\beta_{0}$=0.1 & $\beta_{0}$=1.0 & $\beta_{0}$=10.0\\
\hline
  &     & $M_s \simeq 0.5$  &    \\
  $\theta=0^\circ$     & $0.83\pm0.03$   & $0.88\pm0.02$    & $0.61\pm0.08$   \\
  $\theta=27^\circ$    & $0.96\pm0.01$   & $0.81\pm0.03$    & $0.46\pm0.08$   \\
  $\theta=45^\circ$    & $0.93\pm0.03$   & $0.74\pm0.08$    & $0.41\pm0.13$   \\
  $\theta=63^\circ$    & $0.85\pm0.04$   & $0.48\pm0.13$    & $0.24\pm0.05$   \\
  $\theta=90^\circ$    & $0.13\pm0.07$   & $0.02\pm0.10$    & $0.09\pm0.14$   \\
\hline
  &     & $M_s \simeq 1$    &    \\
  $\theta=0^\circ$     & $0.81\pm0.03$   & $0.82\pm0.02$    & $0.48\pm0.09$   \\
  $\theta=27^\circ$    & $0.92\pm0.02$   & $0.72\pm0.06$    & $0.46\pm0.12$   \\
  $\theta=45^\circ$    & $0.89\pm0.03$   & $0.59\pm0.10$    & $0.48\pm0.12$   \\
  $\theta=63^\circ$    & $0.79\pm0.06$   & $0.42\pm0.09$    & $0.27\pm0.15$   \\
  $\theta=90^\circ$    & $0.00\pm0.08$   & $-0.08\pm0.09$   & $0.07\pm0.08$   \\
\hline
  &     & $M_s \simeq 2$    &    \\
  $\theta=0^\circ$     & $0.73\pm0.03$   & $0.74\pm0.04$    & $0.51\pm0.05$   \\
  $\theta=27^\circ$    & $0.84\pm0.02$   & $0.65\pm0.06$    & $0.43\pm0.08$   \\
  $\theta=45^\circ$    & $0.80\pm0.03$   & $0.59\pm0.09$    & $0.37\pm0.06$   \\
  $\theta=63^\circ$    & $0.73\pm0.07$   & $0.33\pm0.09$    & $0.12\pm0.07$   \\
  $\theta=90^\circ$    & $0.03\pm0.09$   & $-0.04\pm0.18$   & $-0.11\pm0.12$  \\
\hline
\end{tabular}
\caption{Correlation coefficients between RM and $\sqrt{\rm EM}$. The error is the standard deviation of 11 snapshot data.}
\end{center}
\end{table}

\clearpage

\begin{figure}
\vskip -1.2cm
\hskip -1.8cm
\includegraphics[scale=0.82]{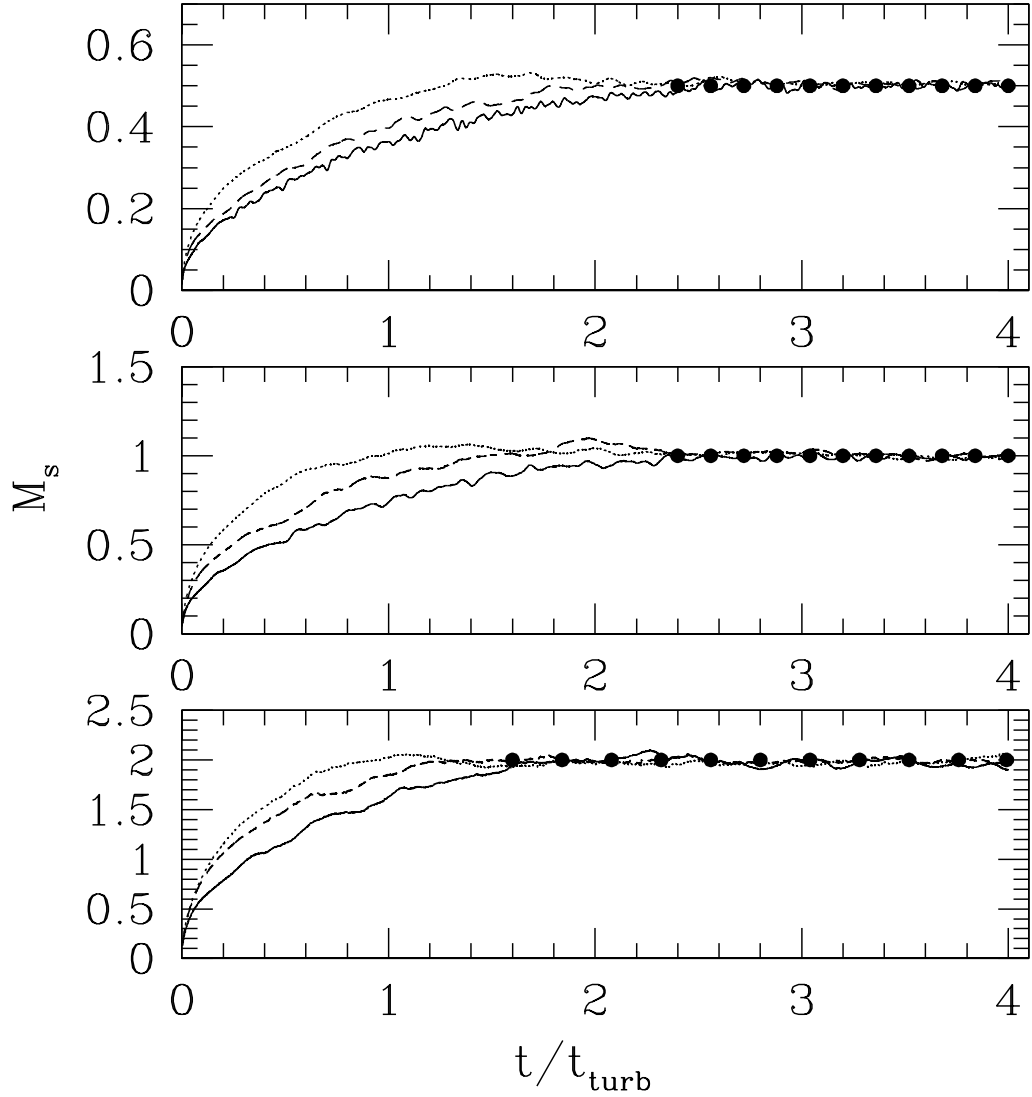}
\vskip -1.5cm
\caption{Evolution of rms Mach number in simulations with $M_s\simeq$ 0.5 (top), 1 (middle), and 2 (bottom) as a function of time in units of $t_{\rm turb} \equiv L/(2 M_s c_s)$. The solid, dashed and dotted lines are for $\beta_{0}$= 0.1, 1 and 10, respectively. Black-dots indicate the epochs of 11 snapshots, whose data were used to calculate the quantities and standard deviations presented in this paper.}
\end{figure}

\clearpage

\begin{figure}
\vskip -2cm
\hskip -0.4cm
\includegraphics[scale=0.72]{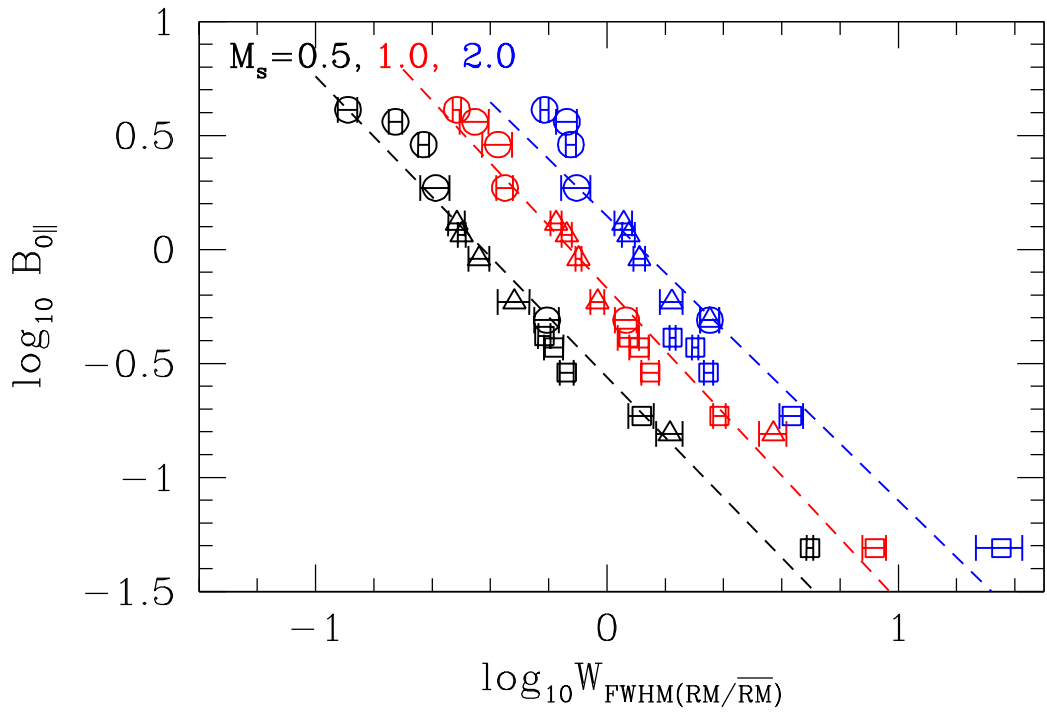}
\vskip -2.5cm
\caption{FWHM of the frequency distribution of RM/$\overline{\rm RM}$ vs. LOS regular magnetic field strength. Black, red, and blue symbols are for $M_s \simeq 0.5$, 1, and 2, respectively. Dashed lines display their best fits. Circles, triangles, and squares are for $\beta_{0}=0.1$, $\beta_{0}=1$, and $\beta_{0}=10$, respectively, for five viewing angles. Error bars show the standard deviation of 11 snapshot data.}
\end{figure}

\clearpage

\begin{figure}
\vskip -2cm
\hskip -0.6cm
\includegraphics[scale=0.74]{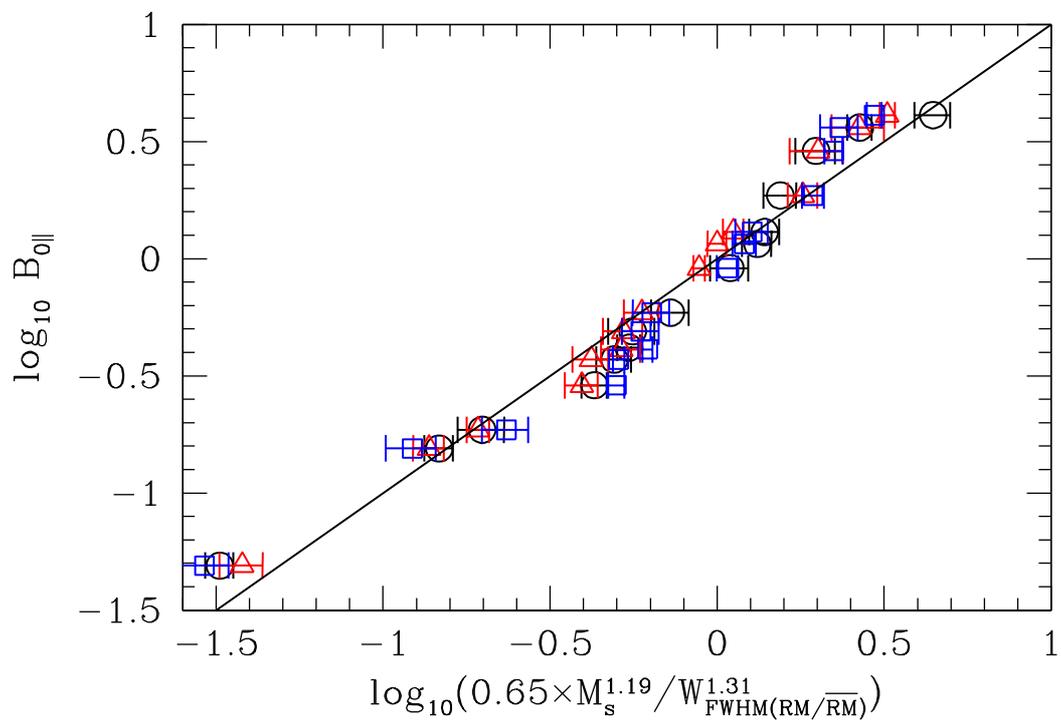}
\vskip -2.5cm
\caption{LOS regular magnetic field strength as a combined function of $W_{\rm FWHM(RM/\overline{RM})}$ and $M_s$. Black circles, red triangles, and blue squares are for $M_s \simeq 0.5$, 1 and 2, respectively. Error bars show the standard deviation of 11 snapshot data. Solid line displays our best fit (Equation (12)).}
\end{figure}

\clearpage

\begin{figure}
\vskip -2cm
\hskip -1cm
\includegraphics[scale=0.76]{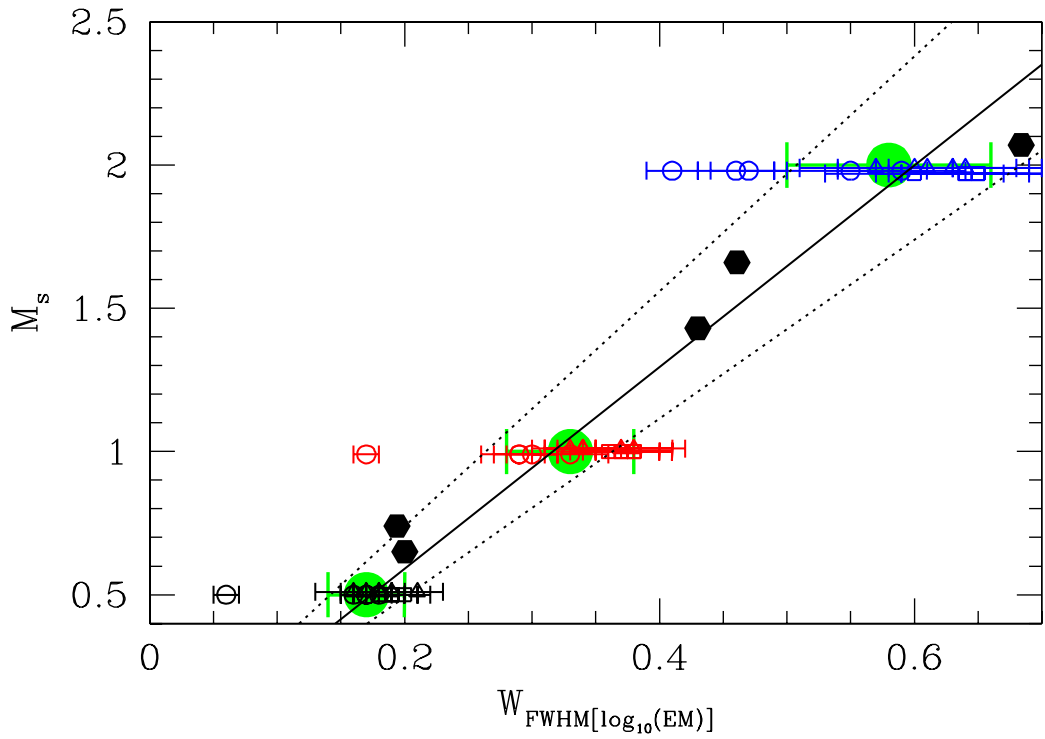}
\vskip -2.5cm
\caption{FWHM of the frequency distribution of $\log_{10} (\rm EM)$ vs. rms Mach number. Open symbols and their error bars are the same as in Figure 2. Filled green circles and their error bars show the average and standard deviation of $W_{\rm FWHM[\log_{10}({\rm EM})]}$ for each $M_s$. Solid and dashed lines display the best fit and 1 $\sigma$ envelope (Equation (13)). Filled black hexagons show the results of \citet{hi08}.}
\end{figure}

\end{document}